\documentclass[preprintnumbers,prd,amssymb,a4paper]{revtex4}

\newcommand{\be}{\begin{equation}}
\newcommand{\ee}{\end{equation}}
\newcommand{\ba}{\begin{eqnarray}}
\newcommand{\ea}{\end{eqnarray}}

\newcommand{\Min}{Minkowski}
\usepackage{color}
\usepackage{latexsym}
\usepackage{amsfonts,amsbsy}

\def\qed{\hbox{${\vcenter{\vbox{
   \hrule height 0.4pt\hbox{\vrule width 0.4pt height 6pt
   \kern5pt\vrule width 0.4pt}\hrule height 0.4pt}}}$}}

\def\onehalf{\textstyle{\frac{1}{2}}}

\def\nablabol{{\stackrel{\circ}{\nabla}}{}}

\begin{document}

\title{Spacetime algebraic skeleton} 

\author{R. Aldrovandi and A. L. Barbosa} 
 \email{analucia@ift.unesp.br} \affiliation{Instituto de F\'{\i}sica Te\'orica --
 Universidade Estadual Paulista \\
Rua Pamplona 145 \\
01405-900 S\~ao Paulo SP \\ Brazil}


\begin{abstract}
 
The  cosmological constant $\Lambda$, which has seemingly dominated the
primaeval Universe evolution and to which  recent data attribute a significant
 present-time value, is shown to have an algebraic
content: it is essentially an eigenvalue of a Casimir invariant of the Lorentz group
which acts on every tangent space. This is found in the context of de Sitter
spacetimes but,  as every spacetime is a 4-manifold with \Min\ tangent spaces,  the
result suggests the existence of a ``skeleton'' algebraic structure underlying
the geometry of general physical spacetimes. Different spacetimes come from the
``fleshening'' of that structure by different tetrad fields. Tetrad fields, which provide
the interface between  spacetime proper and its tangent
 spaces, exhibit to the most the fundamental role of the Lorentz group in
Riemannian spacetimes, a role which is obscured in the more usual metric
formalism.

\end{abstract}

\maketitle




\section{Introduction}

Until rather recently the Universe was believed to follow a
Friedmann solution of Einstein's equations, which comes from using as a source a
gas of matter and radiation~\cite{Wei72,HE73}. The horizon problem forced a
change in this view: at the first stages, the Universe should be
expanding at a much larger rate than Friedmann's prediction, possibly
as it would be if the solution were a de Sitter spacetime. 
Such a spacetime is a solution without any source but with a universal
curvature encapsulated in a cosmological constant.  Its great interest
 comes from de Sitter's finding that Einstein's equations without
any source have such non-flat solutions.  Flat Minkowski space
is a particular case, with vanishing cosmological constant.  The
overall picture would be that of an initial de Sitter Universe which
changes to a Friedmann Universe at an early stage~\cite{Nar93}.  

Recent data point, however, to a significant value for the
cosmological constant even today~\cite{novae1,BBR1}.  This effect is usually simply
added by hand to the Friedmann description.  The new picture is that of a
Universe which begins as a de Sitter spacetime and then changes to the
solution describing the present-day state ---  a Friedmann solution with
a significant de Sitter remnant.  One might wonder whether the
Friedmann ``component'' could not be seen as a perturbation of a
spacetime which remains basically of the de Sitter type.

De Sitter spacetimes have a special characteristic, maximal symmetry. They have ten
Killing vectors, the maximum possible for 4-dimensional spaces. The above
cosmological picture can be rephrased: the Universe is initially empty,  with a
maximal symmetry which is  {\em  partially } broken when matter and radiation somehow
turn up. All this calls for a reappraisal of de Sitter spaces, looking
as deep as possible in its foundations. We intend here to examine the algebraic structure
behind their geometrical make-up. 

Section \ref{sec:roleofLor} sums up the tetrad formalism, emphasis being given to the
manifestations of the Lorentz group --- in particular, to the fact that the
Christoffel-Levi-Civita connection reveals itself as a Lorentz connection when looked at
from any tetrad frame. Section \ref{sec:dS} introduces de Sitter spacetimes through
stereographic coordinates, which suggest a specially simple tetrad frame
in which the subsequent topics are considered. The algebra of
bundle vector fields, in the case a base for the de Sitter Lie algebra, is recalled
in section \ref{sec:vecalg}. The
cosmological constant is examined  through its relation to
curvature  in section \ref{sec:CC}, where it is shown to be ultimately an eigenvalue of a
Casimir invariant operator  of the Lorentz group.

\section{Spacetime: r\^ole of the Lorentz group}
\label{sec:roleofLor}
 
A spacetime $S$ is a four-dimensional differentiable manifold whose
tangent space is, at each point, a Minkowski space.  In bundle
language, the tangent bundle $TS$ on any spacetime has the Minkowski
space $M$ as the typical fiber.  The linkage between the Minkowski
typical fiber and the spaces tangent to spacetime is provided by
tetrad fields.  A tetrad field will solder a copy of $M$ as
tangent space  $T_pS$ at each point $p \in S$.  The tangent bundle is associated to the
principal bundle $BM$ of tetrad frames.  Connections are first defined on $BM$
and then induced on each associated bundle, taking into account the
respective representation of the structure (Lorentz) group.  A spacetime
appears then as the quotient of $BM$ by the Lorentz group. The de Sitter
spacetimes are special because, in their cases, the bundle $BM$ of Lorentzian frames are
Lie groups, just the corresponding de Sitter groups. 

Consider on the typical fiber $M$ the standard canonical vector basis
$\{K_0 = (1,0,0,0), K_1 = (0,1,0,0), K_2 = (0,0,1,0), K_3 =
(0,0,0,1)\}$.  With the convention $a, b, \ldots $ = $0, 1, 2, 3$,
each $K_a$ is given by the entries $(K_a)_b$ = $\delta_{ab}$.  A
tetrad field $ e $ will be a mapping $e: $M$ \rightarrow TS$, $e(K_a)$ =
$e_a$.  This set of four vectors $e_a$ will constitute a local vector
basis on $S$: given a point $p \in S$, and around it an euclidean open
set $U$, the $e_a$'s will constitute a vector basis not only for the
space $T_pS$ tangent to $S$ at $p$ but also for all the $T_qS$ with $q
\in U$.  The extension from $p$ to $U$ is warranted by the
differentiable structure~\cite{AP95}.  The dual forms $\omega^b$, such
that $\omega^b(e_a)$ = $\delta^b_a$, will constitute a covector basis
on the cotangent space at $p$, $T^{*}_pS$.  In most applications the open
set $U$ is a coordinate neighborhood.  Each coordinate system
$\{x^{\mu}\}$ on $U$ will define a ``natural'' (or ``coordinate'')
vector basis $\{\partial_{\mu}$ = $\frac{\partial}{\partial x^{\mu}}
\}$ with its concomitant covector dual basis $\{d x^{\mu}\}$.  This is
a very particular and convenient tetrad field, frequently called a
``trivial'' tetrad, given by $e(K_a)$ = $\delta_a^{\mu}
\partial_{\mu}$.  It is usual to fix a coordinate system around each
point $p$ from the start, and in this sense this basis is indeed
``natural''.  Once such a coordinate system is fixed, with its
accompanying pair of trivial bases, any other tetrad field like the
above generic $\{e_a \}$ and its dual $\{\omega^b \}$ can be written
as~\cite{Cha79,Cha92}
\be %
e_a = h_a{}^{\mu} 
\partial_{\mu} \; \; {\rm and} \; \; \omega^b = h^b{}_{\nu} dx^{\nu} , 
\label{9.1.1} 
\ee %
with the conditions
\be  %
h^b{}_{\mu} h_a{}^{\mu} = \delta^b_a \; \; {\rm and}  \; \; h^a{}_{\mu} 
h_a{}^{\nu} = \delta_{\mu}^{\nu} .
\label{9.1.2}
\ee %
The components $h_a{}^{\mu}(x)$ have one label referring to the
typical Minkowski space and another in tangent spacetime.  We are
using latin letters ($a, b, \dots = 0,1,2,3$) to label components on
$M$ and the greek alphabet for spacetime indices.  We shall call the
first ``Minkowski indices'' (or ``tetrad indices''), the latter
``Riemann indices''.  In their Minkowski labels the tetrads change under
Lorentz transformations according to %
\be h^{ a^{\prime}}{}_{\mu}(x) = \Lambda^{ a^{\prime}}{}_c (x)  \
h^c{}_{\mu}(x) .
\label{9.1.4}
\ee %
$\Lambda^{a^{\prime}}{}_b$ is constant on each tangent space $T_p S$,
but will depend on the point $p$ of spacetime, indicated above by its
coordinates $x = \{x^{\mu}\}$.  Contracting the last expression with
$h_b{}^{\mu}(x)$, we obtain the Lorentz
transformation in terms of the initial and final tetrad basis:  %
\be \Lambda^{ a^{\prime}}{}_b (x) = h^{a^{\prime}}{}_{\mu}(x) 
h_b{}^{\mu}(x) .
\label{9.1.5}
\ee %
Equation (\ref{9.1.4}) says that each tetrad component behaves, on 
each Minkowski fiber, as a Lorentz vector.  More precisely,
$h^a{} = h^a{}_{\mu} dx^\mu$ transforms according to the covector 
representation of the Lorentz group.  A Lorentz covector 
has components changing, under a  transformation with parameters $\alpha^{cd}$,
according to %
\be \phi^{ a^{\prime}} = \Lambda^{ a^{\prime}}{}_b \; \phi^b = 
\left(exp[ \textstyle{\frac{1}{2}} \alpha^{cd} J_{cd}] \right)^{ 
a^{\prime}}{}_b  \; \phi^b .
\label{9.1.6}
\ee %
Here, each $J_{cd}$ is a $4 \times 4$ matrix representing one of the 
Lorentz group generators: %
\be%
 \left[J_{cd}\right]^{a^{\prime}}{}_b = \eta_{db} \;
\delta^{a^{\prime}}_{c} - \eta_{cb} \; \delta^{a^{\prime}}_{d}  \; 
.  \label{9.1.7} 
\ee %
We use the convention $\eta = (\eta_{a b}) = {\rm diag} (1, -1, -1,
-1)$ for the Lorentz metric.  

A tetrad field converts tensors on $M$
into tensors on spacetime.  For example, they will produce a field on
spacetime out of a vector in
Minkowski space by %
\be \phi^{\mu}(x) = h_{a}{}^{\mu}(x) \phi^{a} \; .  \label{9.1.8} \ee 
As the Minkowski indices are contracted, $\phi^{\mu}(x)$ is
Lorentz-invariant.  Another case of tensor transmutation is
$\Lambda^{\mu}{}_{\nu}(x) = h_{a^{\prime}}{}^{\mu} (x) \;
\Lambda^{a^{\prime}}{}_{b}(x) \; h^{b}{}_{\nu}(x) =
\delta^{\mu}_{\nu}$, which shows that there is no Lorentz
transformation on spacetime itself.  The Lorentz metric $\eta_{a b}$ is
transmuted into the Riemannian metric %
\be 
g_{\mu \nu}(x) = \eta_{a b} \; h^{a}{}_{\mu}(x) \; h^{b}{}_{\nu}(x) .  
\label{metfromtet} \ee %
Different tetrad fields transmute $\eta_{a b}$ into different
spacetime metrics.  Of course, also $g_{\mu \nu}(x)$ is
Lorentz-invariant.  The Lorentz group is concealed by the transmutations, and remains
out of sight in the usual metric formalism.  The members of a general tetrad field will
satisfy a set of commutation rules 
\be %
[e_a, e_b] = c^{c}{}_{ab} \; e_{c} .  \label{9.1.11} 
\ee %
The structure coefficients measure its anholonomy and are given by %
 \be %
 c^{c}{}_{ab} = [e_a (h_b{}^{\mu}) - e_b (h_a{}^{\mu})] h^{c}{}_{\mu}
 \; .  \label{9.1.12} 
 \ee %
If $\{e_a \}$ is holonomous ($c^{c}{}_{ab} = 0$), then $\omega^{a}$ =
$dx^a$ for some coordinate system $\{x^a \}$ and $dx^{a^{\prime}}$ =
$\Lambda^{a^{\prime}}{}_b dx^b$.  Expression (\ref{metfromtet}) would
then give just the Lorentz metric written in another system of
coordinates.  In the general Riemannian case, $\{e_a \}$ is
anholonomic.

 A procedure converse to that followed above is more usual: given the
 metric, the tetrad field is defined by (\ref{metfromtet}).  This is
 natural when the metric is known {\it a priori} or, as in General
 Relativity, as the solution of a dynamical equation.  Einstein's
 equations fix the metric and determine the tetrads up to Lorentz
 transformations.

Connections have a special behavior under transmutation.  A general Lorentz connection has
the  form
\begin{equation}
\Gamma = {\textstyle 
\frac{1}{2}} J_{a b}\  \Gamma^{a 
b}{}_{\mu}  \ dx^\mu = {\textstyle 
\frac{1}{2}} J_{a b} \Gamma^{a 
b}{}_{\mu}  \ h_c{}^\mu \omega^c = {\textstyle 
\frac{1}{2}} J_{a b} \Gamma^{a 
b}{}_{c}  \ \omega^c 
\end{equation}
(the factors $1/2$ account for double counting).  The first two indices in the components $\Gamma^a{}_{b \mu}$ are
``algebraic'' and the last a vector index.  Under a Lorentz
transformation $\Lambda^{a^{\prime}}{}_{a}$ = $
h^{a^{\prime}}{}_{\lambda}Õh^{\lambda}{}_{a}$, \be
\Gamma^{a^{\prime}}{}_{b^{\prime} \nu} = \Lambda^{a^{\prime}}{}_{a} \;
\Gamma^{a}{}_{b \nu} \; (\Lambda^{-1})^{b}{}_{b^{\prime}} +
\Lambda^{a^{\prime}}{}_{c} \; \partial_{\nu} \;
(\Lambda^{-1})^{c}{}_{b^{\prime}} .  \label{9.1.13}
\ee %
The Levi-Civita connection is    %
\be %
\Gamma^{\lambda}{}_{\nu \mu} = h_{b}{}^{\lambda} \partial_{\mu}
h^{b}{}_{\nu} + h_a{}^{\lambda} \Gamma^{a}{}_{b \mu} h^{b}{}_{\nu}
\label{9.1.18} 
\ee %
is Lorentz-invariant.  Components $\Gamma^{a}{}_{b \mu} $ and
$\Gamma^{\lambda}{}_{\nu \mu}$ refer to different spaces, but represent he same
connection.  The last expression describes how $\Gamma$ changes when the \Min\ indices are
transmuted into Riemann indices.  Components $\Gamma^{\lambda}{}_{\mu
\nu} $ will, under a base transformation
$B^{\lambda^{\prime}}{}_{\lambda} =
h_{a}{}^{\lambda^{\prime}}h^{a}{}_{\lambda}$, change according to
\[
 \Gamma^{\lambda^{\prime}}{}_{\mu^{\prime} \nu} = 
 B^{\lambda^{\prime}}{}_{\lambda} \Gamma^{\lambda}{}_{\mu \nu} 
 B^{\mu}{}_{\mu^{\prime}} + B^{\lambda^{\prime}}{}_{\lambda} 
 \partial_{\nu} B^{\lambda}{}_{\mu^{\prime}} .  
\] %
A connection acts on a field through the generators of the Lorentz representation it
belongs. On a spinor, for example,  through the spinor generators --- which is the origin of
the name ``spin connection'' frequently given to $\Gamma^a{}_{b \mu}$. 
On a vector $U$, it acts  with the
generators of Eq.(\ref{9.1.7}):
$$%
\Gamma^a{}_{c \mu} U^{c}  = \onehalf\ \Gamma^{ef}{}_{\mu} (J_{ef})^a{}_{c} U^c . 
$$%
A good illustration of this action turns up when we say that a vector ``stays the same'' by parallel
transport. Consider, for example, along a curve with parameter
$u$, the covariant derivative of its tangent velocity $U$:
$$%
\frac{\nabla U_\lambda}{\nabla u} = \frac{d U_\lambda}{d u} - \Gamma^{\mu}{}_{\lambda \nu} U_{\mu} U^{\nu}
=  \frac{d U_\lambda}{d u} +  U_\nu \partial_\lambda U^\nu = U^\nu (\partial_\nu U_\lambda  - 
\partial_\lambda U_\nu). 
$$%
Seen from the tetrad frame $e_a = h_a{}^\mu \partial_\mu$, this leads to
$$%
\frac{\nabla U_a}{\nabla u} = \frac{\nabla h_a{}^\lambda U_\lambda}{\nabla u} 
= U_c\ h^c{}_\lambda \frac{\nablabol h_a{}^\lambda }{\nabla u} 
+ U^b [e_b ( U_a) - e_a ( U_b)  +c^{e}{}_{ab} U_e] . 
$$%
As $h^c{}_\lambda \frac{\nablabol h_a{}^\lambda }{\nabla u} =  \Gamma^{c}{}_{ab} U^b$ and 
$\Gamma_{cab} U^b U^c $ = $ -\ \onehalf\ [c_{cab} + c_{abc} + c_{bac} ] U^b U^c $ = $ -\ c_{cab} U^b U^c$,
$$%
 \frac{\nabla U_a}{\nabla u} = U^b [\delta^c_{b} \delta^d_a - \delta^c_{a} \delta^d_b] e_c (U_d)  +
(c^{e}{}_{ab} +
\Gamma^{e}{}_{ab} ) U_e] =  U^b (J_{ba})^{cd} e_c (U_d) .
$$%
Along a general curve, the velocity $U$ is continuosly changed by
infinitesimal Lorentz  transformations.   These transformations vanish along a geodesic.

\section{de Sitter spacetimes}
\label{sec:dS}
There are only two kinds of de Sitter spacetimes~\cite{HE73}: that
usually called de Sitter spacetime
proper, and the so-called anti-de Sitter spacetime.  
Their respective groups of motions are the two de Sitter
groups $SO(4,1)$ and $SO(3,2)$.  If the
curvature is made to tend to zero, both groups reduce to the Poincar\'e group 
$SO(3,1) \oslash T^{3,1}$ by a
Inon\"u-Wigner contraction~\cite{gursey} and both spaces reduce to \Min\ space.

These three groups have much in common~\cite{AP86,AP88}: their
manifolds are fiber bundles structured by the Lorentz group ($SO(3,1)$ or its covering). 
Base spaces are the respective spacetimes.  Thus, the Poincar\'e group is
the complete space of a principal bundle with Minkowski space as base
space, and each de Sitter group is the complete space of a principal
bundle with the corresponding de Sitter space as base space.  The
group manifolds coincide, in these cases, with the bundles of
Lorentzian frames on the respective spacetimes. The latter are the quotients
$SO(4,1)/SO(3,1)$, $SO(3,2)/SO(3,1)$  and $(SO(3,1) \oslash T^{3,1})/SO(3,1)$.
  
We shall introduce the de Sitter and the anti-de Sitter spacetimes as hypersurfaces in
the pseudo--Euclidean spacetimes ${\bf E}^{4,1}$ and ${\bf E}^{3,2}$,
inclusions whose points in Cartesian coordinates $(\xi) = (\xi^0,
\xi^1, \xi^2, \xi^3$, $\xi^4)$ satisfy, respectively, 
\ba %
 (\xi^0)^2 - (\xi^1)^2 - (\xi^2)^2 - (\xi^3)^2  - (\xi^4)^2 &=\eta_{a b} \, \xi^{a}
\xi^{b}  - (\xi^4)^2 &=-\ L^2 \; ; \nonumber
\\
 (\xi^0)^2 - (\xi^1)^2 - (\xi^2)^2 - (\xi^3)^2  + (\xi^4)^2 &= \eta_{a b} \, \xi^{a}
\xi^{b}  + (\xi^4)^2 &= \ \ \ L^2 \; .  \nonumber 
\ea %
$L$ is a constant ``pseudo-radius''. The sign notation
$\eta_{44} = s$ allows these conditions to be put together as
\[
 s\ \eta_{a b} \, \xi^{a} \xi^{b} + 
\left(\xi^4\right)^2 =
L^2 \; .
\]
We now change to stereographic coordinates $\{x^\mu\}$ in
4--dimensional space, which are given by 
\be %
x^\mu = h_{a}{}^{\mu} \,
\xi^{a} \; ,
\label{41}
\ee %
where a specially simple tetrad field turns up:
\be %
h_{a}{}^{\mu} = {\textstyle \frac{1}{n}} \,
\delta_{a}^{\mu} \; , {\ \ \textnormal {where}\ \ } n = \onehalf
\left(1 - \frac{\xi^4}{L} \right) = \frac{1}{1+ s \sigma^2/4L^2} {\ \
\textnormal {and}\ \ } \sigma^2 = \eta_{a b} \, \delta^{a}{}_{\mu}
\delta^{b}{}_{\nu} \, x^\mu x^\nu .
\label{42}
\ee %
Calculating the line element on the hypersurfaces, we find
$ds^2 = g_{\mu \nu} \,d x^\mu dx^\nu$, where 
\be %
g_{\mu \nu} = h^{a}{}_{\mu} h^{b}{}_{\nu} \eta_{a b} = n^2
\delta^{a}_{\mu} \delta^{b}_{\nu} \eta_{a b} \ .
\label{44}
\ee %
The most natural tetrad base and its dual are 
\be%
e_{a} = h_{a}{}^{\mu} \partial_\mu = {\textstyle \frac{1}{n}}\
\delta_{a}{}^{\mu} \partial_\mu \; \ \ \ {\textnormal {and}} \ \ \
\omega^{a} = h^{a}{}_{\mu} dx^\mu = n \delta^{a}{}_{\mu} dx^\mu \; . 
\label{duba}
\ee%
Equation (\ref{44}) tells us that de Sitter spacetimes are conformally
flat~\cite{Wei72} with conformal factor $n^2(x)$.  The Christoffel
symbols corresponding to the above metric are
\be%
\Gamma^{\lambda}{}_{\mu \nu} = \left[ \delta^{\lambda}{}_{\mu}
\delta^{\sigma}{}_{\nu} + \delta^{\lambda}{}_{\nu}
\delta^{\sigma}{}_{\mu} - g_{\mu \nu} g^{\lambda \sigma} \right]
\partial_\sigma \ln n \; ,
\label{46}
\ee%
leading to the Riemann and Ricci tensor components
\be%
R^{\alpha}{}_{\beta \rho \sigma} = -\ \frac{s}{L^2} 
\left[\delta^{\alpha}_{\sigma} g_{\beta \rho} - \delta^{\alpha}_{\rho} 
g_{\beta \sigma} \right] , \ \ \  R_{\mu \nu} \ = \frac{3 s}{L^2} \ g_{\mu \nu} ,
\label{47}
\ee%
and to the scalar curvature 
\begin{equation}%
R = 12 \ \frac{s}{L^2} \ . \label{deSitterRscalar}
\end{equation} %

\section{The algebra of vector fields}
\label{sec:vecalg}
Distinct de Sitter spacetimes (including {\Min}) differ by their
curvatures, and bundles are the natural place to study connections and
their curvatures. The local geometrical properties are specially evident in the space
tangent to the bundle.  As the groups coincide with the bundles, the
vertical tangent vector fields can be directly related to the
generators.  Thus, from a local point of view, that difference in
curvature should be reflected in the corresponding Lie algebras. Besides the Lorentz generators
$\{J_{ab}\}$ of Eq.(\ref{9.1.7}),  we shall use the generators $\{T_c\}$ of spacetime
translations.

Given the Lorentz metric $\eta_{a b}$, consider the 
three sets of structure constants:
\be %
f^{(e f)}{}_{(a b)(c d)} = \eta_{b c}
\delta^{e}{}_{a} \delta^{f}{}_{d} + \eta_{a d}
\delta^{e}{}_{b} \delta^{f}{}_{c} - \eta_{b d}
\delta^{e}{}_{a} \delta^{f}{}_{c} - \eta_{a c}
\delta^{e}{}_{b} \delta^{f}{}_{d} \; 
\label{LL}
\ee%
\be%
f^{(e)}{}_{(a b)(c)} = \eta_{c b} \delta^{e}{}_{a} - \eta_{c a}
\delta^{e}{}_{b} \;
\label{LT}
\ee%
\be%
f^{(e f)}{}_{(a)(b)} = \left(\delta^{f}{}_{a} \delta^{e}{}_{b}
- 
\delta^{e}{}_{a}
\delta^{f}{}_{b} \right) = -\ f^e{}_{(ab)}{}^f\; .
\label{TT}
\ee
The Lie algebras of the de Sitter groups are then given  by
\ba %
\left[J_{c d}, J_{e f}\right] &=& {\textstyle \frac{1}{2}} f^{(a
b)}{}_{(c d)(e f)} \, J_{a b} \label{dSLL} \\
\left[J_{c d},T_{e}\right] &=& f^{(a)}{}_{(c
d)(e)} \, T_a   \label{dSLT} \\
\left[T_c, T_e \right] &=& {\textstyle 
\frac{1}{2}} \, s \ f^{(a  b)}{}_{(c)(e)} \, J_{a b} \; .
\label{dSTT}
 \ea %
The difference with respect to the
Poincar\'e group lies in the
 non-commutativity of de Sitter translations.  We can use Eq.(\ref{dSTT}) all
 the way long, putting $s = 0$ in the Poincar\'e case. Notice the dimensional anomaly of this equation:
translation generators have dimension length$^{-1}$, while Lorentz generators are dimensionless. Though 
acceptable as an abstract relation, it must be corrected in any physical application. This has been
done by the factors of $L^2$ which turned up in the previous section.

The direct-product character of the bundle is recovered by using a tetrad to
take the translations ``down'' to spacetime. We shall have $\partial_\mu$ or
$\nabla_\mu = h^a{}_\mu T_a$, depending on whether we are interested in a
common derivative or a covariant one. In both cases, the middle commutators (\ref{dSLT})
vanish. For example, $[J_{cd},
\partial_\mu] = 0$ because $h^a{}_\mu$, with a covector index $a$, transforms just in the opposite way to 
$T_a$, for which $a$ is a vector index:  $J_{cd}(h^a{}_\mu) = -\ f^{(a)}{}_{(c d)(e)} h^e{}_\mu$.

It can
 be verified from (\ref{LT}) that the Lorentz metric satifies
\be%
\eta_{d e} \, f^{(e)}{}_{(a b)(c)} + \eta_{c e} \, f^{(e)}{}_{(a
b)(d)} = 0 \; .
\label{Lor3}
\ee  %
This provides a precise meaning to raising and lowering indices of the
structure constants and their derived quantities.  If we accept to
lower the entry indices with the Lorentz metric, then the matrices
$f_{(a b)} $ with entries $ (f_{(a b)})_{e c}$ = $f_{(e)(a b)(c)} $
are antisymmetric.  Applying tetrads we find that also
$g_{\mu \nu} \, f^{(\nu)}{}_{(a b)(\lambda)} + g_{\lambda \nu} \,
f^{(\nu)}{}_{(a b)(\mu)} = 0$.
Metrics satisfying this type of condition, the same as Eq.(\ref{Lor3}), are invariant
metrics of the algebra~\cite{KN63}.

 Additional information can be obtained from the Jacobi identities. 
 There are four types, according to the numbers of generators of each
 kind involved: $[L,[L,L]]$, $[L,[L,T]]$, $[L,[T,T]]$ and $[T,[T,T]]$. 
 Some of them give no information at all in the $s = 0$ Poincar\'e case. 
The Lorentz sector constitutes a closed
subalgebra, so that type
 $[L,[L,L]]$ has nothing to say on the translations.  The $[L,[L,T]]$
 identity
$$%
 \left[ J_{a b},\left[J_{c d},T_{e}\right] \right] + 
 \left[T_{e},\left[J_{a b},J_{c d} \right] \right] +
 \left[J_{c d},\left[T_{e},J_{a b} \right] 
 \right]  = 0
 $$%
 gives the condition
$$%
 f^{(h)}{}_{(a b)(f)} f^{(f)}{}_{(c 
 d)(e)} -  
 f^{(h)}{}_{(c d)(f)}f^{(f)}{}_{(a 
 b)(e)} = \; {\textstyle \frac{1}{2}} f^{(
 f)}{}_{(a b)(c d)} f^{(h)}{}_{(g
 f)(e)} .
$$%
Comparison with Eq.(\ref{dSLL}) tells that each generator $J_{a b}$ can be
represented by a $4 \times 4$ matrix with elements
\be %
\left( J_{a b} \right)^{c}{}_{e} = f^{(c)}{}_{(a b)(e)}.
\ee%
This corresponds precisely to the vector representation (\ref{9.1.7})
of the Lorentz group and gives the meaning of the second lines in the
commutation tables above: the translation generators are Lorentzian
vectors. By Eq.(\ref{Lor3}), the lowered-index matrices are
antisymmetric. 

\section{The cosmological constant}
\label{sec:CC}

The action of $J_{a b}$ on the vectors $T_c$ is summed up in 
Eq.(\ref{dSLT}): through the commutator in the left-hand side, by
matrix multiplication in the right-hand side.  
This vector representation will have an important role in what follows.  Indeed,
the $[T,[T,T]]$ Jacobi identity gives 
\[
s\ f^{(h)}{}_{(e 
 f) [ (a )}  f^{(e 
 f)}{}_{(b)(c) ]} = 0,
\]
where $[a b c]$ stands for the summation over all cyclic permutations
of the included indices.  The left factor is a matrix element $\left(
J_{e f} \right)^{e}{}_{a}$ for a Lorentz generator in the vector
representation.  We can interpret each right factor as a
component of some matrix along $ J_{e f}$: introduce the
matrices $F_{b c}$ with dimension of curvature ($L^{-2}$) and
components $f^{(e f)}{}_{(b)(c)}$ along each $ J_{e f}$:
\be %
F_{a b} = {\textstyle \frac{s}{2L^2}} J_{e f} f^{(e f)}{}_{(a)(b)}
. 
\ee %
Such matrices will have entries $ R^{c}{}_{d a b} = (F_{ab})^{c}{}_{d}$, or %
\be %
R^{c}{}_{d a b} = {\textstyle \frac{s}{2L^2}} f^{(c)}{}_{(ef) (d
)} f^{(ef)}{}_{(a)(b) } . \label{Reabc}
\ee%
Combined with metric invariance (\ref{Lor3}), the antisymmetry of
$R^{e}{}_{a b c}$ in the first two indices is obtained.  The
$[T,[T,T]]$ Jacobi identity will then say that
\be%
 R^{c}{}_{[ d a b ]} = 0.   \label{cyclicident}
\ee%
This is a suggestive expression, formally identical to the
vanishing--torsion Bianchi identity (``cyclic property'') for the Riemann tensor.  This
means that the Bianchi identity is the geometrical version of the
Jacobi identity, or the Jacobi identity is the algebraic version of
the Bianchi identity.

Adding a connection will correspond to a non-trivial
extension~\cite{Ald91,AB00} of the translation algebra, that is, from
 $s = 0$ to $s \ne 0$ in Eq.(\ref{dSTT}).  We should perhaps
change the notation acoording to $T_a \Rightarrow D_a$, as the ``extended''
translation is actually a covariant derivative
\begin{equation}%
D_c =  e_c + \onehalf A^{ab}{}_c J_{ab}. \label{Dc}
\end{equation}%
The $A^{ab}{}_c$'s are the connection components in base $\{h_a \}$
and the $J_{ab}$'s are the generators in the corresponding
representation. Applied to a vector field, for example, their commutator is, by
Eq.(\ref{dSTT}),
\begin{equation}%
[D_a, D_b] V^e =  \onehalf s f^{(cd)}{}_{(a)(b)} (J_{cd})^e{}_f V^f = 
\onehalf s f^{(cd)}{}_{(a)(b)} f^{(e)}{}_{(cd)(f)} V^f .\label{DaDb}
\end{equation}%
 Notice the difference in the meaning of the indices: the
pair (ab) in $A^{ab}{}_c J_{ab}$ indicates the component of matrix $A$
along $J_{ab}$. The usual indices in the standard notation signify matrix
entries, $A^{ab}{}_c = \onehalf A^{cd}{}_c (J_{cd})^{ab}$. Our notation has
been chosen to make them coincide. Comparison of (\ref{DaDb}) with the standard
formula  
$$%
V^{c}{}_{;a;b} - V^{c}{}_{;b;a} = R^{c}{}_{d a b} V^{d}
$$%
shows that $R^{c}{}_{d a b}$ as introduced in 
Eq.(\ref{Reabc}) is indeed the curvature tensor.  We have then a strong
relationship with the de Sitter algebra. Using Eqs.(\ref{TT}) and (\ref{Lor3}), we
find that the Ricci tensor is a matrix element:
\begin{equation}%
R_{a b} = -\ {\textstyle \frac{s}{2L^2}}  
f_{(b)}{}^{(cd)}{}_{(e)} f^{(e)}{}_{(cd)(a)} = -\ 
{\textstyle \frac{s}{L^2}} (J^2)_{ab}
\end{equation}%
 with
$J^2 =\onehalf J^{(ab)} J_{(ab)}$ a Casimir operator of the Lorentz
group.  The scalar curvature is consequently its trace:
\begin{equation}%
R = -\ {\textstyle \frac{s}{L^2}} {\textnormal {tr}} J^2 .
\end{equation}%
As $\left[J^2\right]^a{}_{b} = -\ 3  \delta^a{}_{b}$ from Eq.(\ref{TT}), we recover indeed the
``geometrical'' expression of  Eq.(\ref{deSitterRscalar}). On the other hand Einstein's equation for the
case,
\begin{equation}%
R^a{}_{b} - \onehalf \delta^a{}_{b} R = \Lambda \delta^a{}_{b} ,\label{Eeq}
\end{equation}%
acquires an algebraic version:
\begin{equation}%
-\ {\textstyle \frac{s}{L^2}} \left[J^2 -  \onehalf ({\textnormal {tr}} J^2) I
\right] = \Lambda \ I .
\end{equation}%
We find from Eq.(\ref{Eeq}) that the cosmological constant is $\Lambda = - R/4$, or
\begin{equation}
\Lambda = {\textstyle \frac{s}{4 L^2}} {\textnormal {tr}} J^2 = {\textstyle \frac{s}{ L^2}}\
\times
{\textnormal {the eigenvalue of}}\  J^2 \ {\textnormal {in the vector
representation.}}
\end{equation}
It is, consequently, a purely algebraic invariant of the Lorentz group. Notice that it is the
numerical value of the parameter $L$ which is measured experimentally. De Sitter spaces
are, of course, unique both because of their conformal character and because their Lorentz
bundle is the manifold of a group, which is furthermore semisimple. They have one further exceptional
feature: as symmetric homogeneous spaces, their curvatures satisfy the Yang-Mills
equations~\cite{NT78,HTS80}. They stand nearer to
the ``skeleton'' structure than any other spacetime. Other spacetimes ---  the Friedmann model, for
example -- will differ in two aspects: they will not be conformally flat, and their Lorentz
bundles will have  less symmetry.

Let us finish, however, by recalling a remarkable fact. In the favored Friedmann model with a
cosmological constant, the Universe spends most of its lifetime in a radiation-dominated stage. It so
happens that radiation does not contribute to the scalar curvature. The  relation $ \Lambda = -
R/4 $  and the above ``algebraic'' result consequently keep holding.

\section{Summing up}

Recent observation data point  to a significant
present-day value for the cosmological constant $\Lambda$.  Added to the necessity of
$\Lambda$-dominance at the earliest stages to provide for
inflation, this leads to a picture of a Universe which starts as a de
Sitter spacetime and is subsequently deformed by the effect of radiation and matter.
Any spacetime is a differentiable manifold every point of which has for tangent
space a Minkowski space, on which the Lorentz group acts. The whole structure is summed up
in the bundle of Lorentzian frames on spacetime.
 On that bundle, the Lorentz generators
are represented by vertical vector fields and the Minkowski  vectors by  horizontal
fields. An analysis of the complete algebra of vector
fields on the bundle shows that
$\Lambda$ is the eigenvalue of a Casimir operator of 
 the Lorentz group. 

\begin{acknowledgments}
The authors are thankful to FAPESP and CNPq, brazilian agencies, for financial support.
\end{acknowledgments}

\end{document}